\documentclass[aip,jcp,reprint]{revtex4-1}
\usepackage{amsmath} 
\usepackage{esint}
\usepackage{graphicx}
\usepackage{color}

\begin{document}
\title{Sound Attenuation in Glasses} 
\author{Grzegorz Szamel}
\author{Elijah Flenner}
\affiliation{Department of Chemistry, Colorado State University,
Fort Collins, Colorado 80523, USA}
\email[The authors to whom correspondence may be addressed: ]{grzegorz.szamel@colostate.edu, flennere@gmail.com}

\date{\today}

\begin{abstract}
 Comprehending sound attenuation is integral to understanding the anomalous low temperature properties of glasses. Despite decades of studies, 
 the underlying mechanism of sound attenuation in glasses is still debated. In this perspective
 we review recent work on sound attenuation in amorphous solids. We focus on the role of defects, heterogeneous elasticity, and we also discuss attenuation in model amorphous solids without
 defects. We review our definition of attenuation defects and show that they strongly influence sound attenuation. However, we also find another contribution to sound attenuation that cannot
 be attributed to attenuation defects. We confirm an earlier result of Kapteijns \textit{et al.} [G. Kapteijns \textit{et al.},  J. Chem. Phys. \textbf{154}, 081101 (2021)]
 that heterogeneous elasticity theory predicts relative changes of sound attenuation in model two-dimensional glasses if the configuration-to-configuration elastic constants fluctuations
 are used to quantify the heterogeneity. We extend this finding to similar three-dimensional glasses. We end by discussing the Euclidean Random Matrix model, which exhibits 
 Rayleigh scaling of sound attenuation, but does not have quasi-localized excitations, and thus probably does not have sound attenuation defects. We propose that the mechanisms
 behind sound attenuation can be more fully understood by approaching the problem from two directions, one where the strong influence of defects is studied and another where
 sound attenuation is studied in defect free albeit disordered materials.  
\end{abstract}

\maketitle

\section{Introduction}

From a macroscopic point of view, ignoring the microscopic structure, both crystalline solids and amorphous solids can be considered elastic bodies, and for this reason  
it was originally expected that the low-energy excitations in both classes of materials  are plane waves.  
Debye theory, which assumes that the low-energy excitations are plane
waves, makes reasonable predictions for the low temperature properties of crystalline solids. 
However, about 50 years ago if was found that Debye theory does not predict the low temperature
properties of glasses \cite{Zeller1971,Ramos2023}. 
The disordered, microscopic structure of the glass influences the behavior of the glass at a macroscopic level. It was also found
that there are more vibrational modes than predicted by Debye theory \cite{Buchenau1984,Schroeder2004}. These excess modes results in a peak in the density of states divided by
the Debye theory prediction, \textit{i.e.}~in the reduced density of states. The influence of the disordered
structure can also be clearly observed in the much stronger attenuation of sound in glasses compared to crystals \cite{Ramos2023,Yu1987,Zaitlin1975,Buchenau1984}.   

The strength of the sound attenuation in glasses is striking when one compares the time dependence of standing sound waves
in classical simulated crystals and glasses in the harmonic approximation 
\cite{GelinNatMat2016,Wang2019a,Mizuno2017,Kapteijns2021,Mahajan2021}.  One method to study sound attenuation is to excite a standing
plane wave 
and examine the decay of velocity correlations \cite{GelinNatMat2016,Mizuno2017,Wang2019a}.  For simulated crystals, the
velocity correlation function does not decay since the plane wave is an eigenvector of the Hessian. In contrast, for glasses the velocity correlation function decays since
the plane wave is no longer an eigenvector of the Hessian. The energy of the excited plane wave is transferred into harmonic vibrational modes, given by the eigenvectors of the 
glass's Hessian, exciting modes with frequencies around the frequency of the sound wave \cite{Moriel2019}.

While this mathematical description explains what occurs in simulations, it is of little help in understanding the physical processes leading to sound attenuation in experiments.
In experimental systems one needs to not only understand sound attenuation in the classical harmonic approximation, but also quantum and finite temperature contributions.  
At temperatures below one Kelvin, quantum two-level systems are known to dominate sound attenuation. The seminal work of Zaitlin and Anderson\cite{Zaitlin1975} 
demonstrated that sound waves 
control the thermal conductivity of dielectric glasses, and that the universal nature of thermal conductivity can be attributed to two-level systems. When the temperature 
is increased to above 1K there is a plateau in the thermal conductivity, which can be rationalized by postulating a Rayleigh scattering contribution to sound attenuation 
\cite{Ramos2023}. 

The source of a Rayleigh scattering contribution has been discussed in many theoretical works on sound attenuation.
An early idea, the soft-potential model, made a connection between the quantum two-level systems and Rayleigh scattering scaling in three dimensional systems
\cite{Ramos2023,Buchenau1972,Schober2011}. 
The soft potential model assumes the existence of quasi-localized excitations which are responsible for two-level systems but also couple to sound waves.
At low temperatures, the soft potential model predicts an additional 
$\omega^4$ contribution to the vibrational density of states in addition to the contribution due to Debye theory, which scales as $\omega^{d-1}$, where $d$ is the dimensionality. The coupling of the
quasi-localized excitations with the sound waves results in an $\omega^4$ scaling of sound attenuation. The soft-potential model also includes the 
contribution from sound attenuation below 1K due to quantum two-level systems, and thus it can describe sound attenuation over a wide range of temperatures. However,
it has been challenging to directly verify the theory in simulations \cite{Shimada2018,Buchenau2021}.  Nevertheless, the connection between quasi-localized excitations and
sound attenuation is a common, returning theme in the literature. 

Another prominent theory that predicts Rayleigh scattering scaling low frequency sound attenuation is fluctuating elasticity theory
(FET) \cite{Schirmacher2010,Schirmacher2011,Marruzzo2013,Ganter2010}. 
When the wavelength of sound is much larger than the spatial scale of the local elasticity fluctuations,
FET predicts Rayleigh scattering scaling of sound attenuation. Therefore, the soft potential model and 
FET predict an $\omega^4$ sound attenuation scaling in three dimensions, but FET predicts $\omega^3$ scaling in two dimensions.
Interestingly, FET also predicts that there is a relationship between sound attenuation and the boson 
peak, with the excess vibrational modes growing with frequency with the same power law as sound attenuation \cite{Marruzzo2013}. 

While the soft potential model and the fluctuating elasticity theory provide a physical picture of the mechanism of sound attenuation, ideally one would like to derive the 
parameters used in these theories from first principles or from computer simulations. Some alternative recent theories have focused on the role of the non-affine forces that arise in glasses
due to a macroscopic deformation. These forces result in non-affine displacements that do not occur in crystals after a macroscopic deformation \cite{Baggioli2022,Szamel2022}. 
We have recently shown that sound attenuation in the harmonic approximation can be calculated from the properties of the non-affine forces and the eigenvectors and eigenvalues of the Hessian 
calculated at an inherent structure \cite{Szamel2022}. Notably, our approach does not involve any adjustable parameters. 
However, our theory does not rule out a physical picture of quasi-localized excitations or fluctuating elasticity. 

In this perspective we examine simulational and theoretical results on sound attenuation in glasses, focusing on results in the harmonic approximation or at $T=0$. We take a broad look at the 
sound attenuation in glasses and provide an overview on how some approaches may be interestingly related.  We limit ourselves to a comparison with computer simulations since
our theory \cite{Szamel2022} needs eigenvalues and eigenvectors of the Hessian matrix in order to evaluate sound attenuation coefficients. 

\section{Sound attenuation, Non-Affine Forces, and Vibrational Modes}\label{nonaff}

\begin{figure}
  \includegraphics[width=0.95\columnwidth]{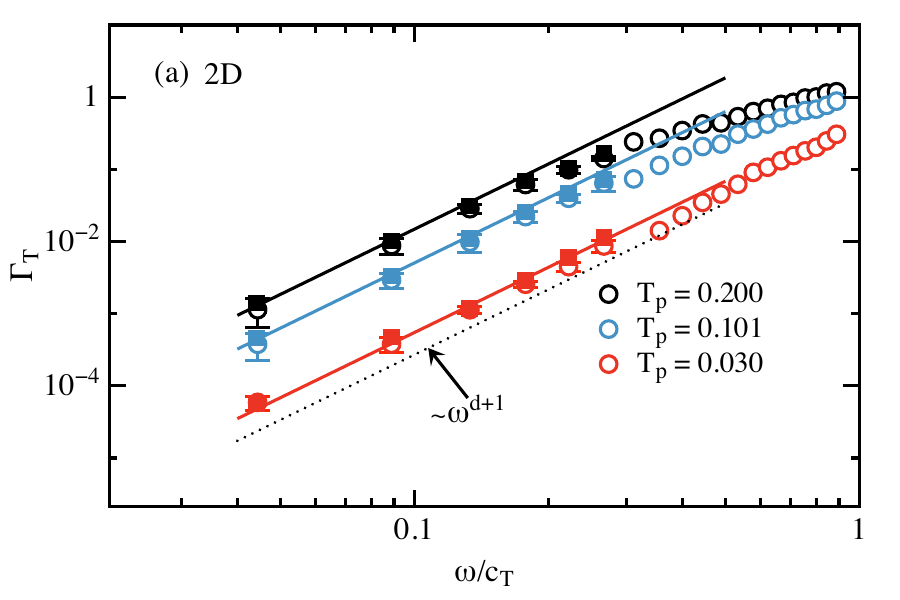}
  \includegraphics[width=0.95\columnwidth]{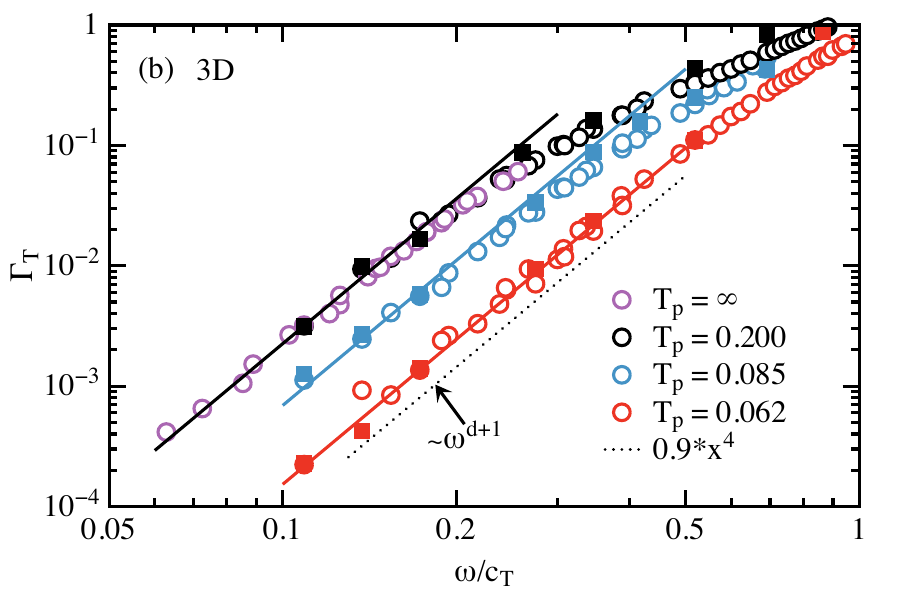}
  \caption{\label{fig:attentheory} Comparison of sound attenuation coefficients predicted by the theory with coefficients calculated from
    direct simulations.  Filled squares: full theory of Ref. \citenum{Szamel2022}; lines: small wavevector limit of the full theory, Eq.~\eqref{eq:damp};
    open circles: results of direct sound attenuation simulations \cite{Wang2019a,Flenner2024}. $T_\text{p}$ is the parent temperature of the glass \cite{Wang2019a,Wang2019};
  glass stability increases with decreasing parent temperature. Dotted lines show Rayleigh scattering scaling, $\Gamma\propto\omega^{d+1}$.}
 \end{figure}

In a series of works we recently examined the relationship between sound attenuation, the non-affine force field, and harmonic vibrational modes \cite{Szamel2022,Flenner2024,Flenner2024b}. 
These works stemmed from
a theory of low-frequency sound attenuation in the harmonic approximation developed by Szamel and Flenner \cite{Szamel2022}.
In the harmonic approximation sound propagation and attenuation are governed by the matrix of the second derivatives of the potential energy, \textit{i.e.} the Hessian,
calculated at a potential energy minimum (inherent structure),
\begin{eqnarray}\label{eq:H}
\mathcal{H}_{ij}^{\alpha\beta} = \sum_{l\neq i}
\frac{\partial^2 V(R_{il})}{\partial R_i^\alpha \partial R_j^\beta} 
\end{eqnarray}
where $V(r)$ is the pair potential, $R_i^{\alpha}$ is the $\alpha$ direction of the vector $\mathbf{R}_i$, and 
$\mathbf{R}_i$ is the inherent structure position of particle $i$. 
In the small wavevector limit, relevant for low frequencies, 
we showed that sound attenuation can be calculated from
\begin{equation}\label{eq:damp}
\Gamma_a(\omega) = \frac{\omega^2}{v_a^2} \sum_{\omega_p} \delta(\omega - \omega_p) \epsilon(\omega_p),
\end{equation}
where $\omega_p$ are the frequencies corresponding to the eigenvectors of the Hessian, and $v_a$ is the speed of sound for a transverse wave, $a = T$, 
or a longitudinal wave, $a = L$. In Eq.~\eqref{eq:damp} $\epsilon(\omega_p)$ is the projection of the non-affine force field, $\mathbf{\Xi}_a$, 
 that originates from a transverse or longitudinal deformation onto an eigenvector $\boldsymbol{\mathcal{E}}(\omega_p)$
 of the Hessian with eigenvalue $\omega_p^2$,
 \begin{equation}\label{eq:eps}
   \epsilon(\omega_p) = (\pi)/(2 \omega_p^2)
   \frac{1}{N} \sum_i \left| \mathbf{\Xi}_{a,i}\cdot \boldsymbol{\mathcal{E}}_i(\omega_p)\right|^2,
   \end{equation}
 where $\mathbf{\Xi}_{a,i}$ and $\boldsymbol{\mathcal{E}}_i$ are components of the non-affine force field and the eigenvector of the
 Hessian pertaining to particle $i$, respectively.
 The deformation that determines the non-affine force field is a simple shear if 
 one is calculating sound attenuation for transverse waves or a compression if one is calculating sound attenuation for longitudinal waves. There are
 different geometrical directions of simple shear deformations and compressions, and to get the best statistics one averages over them.  We emphasize that 
 Eq.~\ref{eq:damp} has no adjustable parameters and can be calculated from knowledge of the interaction potential. Therefore, it can be considered a first principles 
 theory of sound attenuation, but it does not exclude the validity of other descriptions.
 The predictions of our theory agree very well with direct simulations of sound attenuation, see Fig. \ref{fig:attentheory}.
 In this section we will examine some implications of the theoretical expression for sound attenuation, Eq.~\eqref{eq:damp}.
 
 We begin by examining the non-affine force field $\mathbf{\Xi}_a$, which depends on the precise nature of the deformation, and we focus on a shear deformation.
 The deformation is specified by an applied strain $\epsilon_{\alpha \beta}$ and thus the subscript $a$ is now replaced by $\alpha\beta$. 
 The non-affine force field can be calculated from the Hessian matrix $\mathcal{H}_{ij}^{\gamma\delta}$. Specifically, for 
 an applied strain $\epsilon_{\alpha \beta}$ the non-affine force on particle $i$ in the direction $\kappa$ is given by
 \begin{equation}\label{eq:naforce}
 \Xi_{\alpha \beta,i}^\kappa = -\frac{1}{2} \sum_j \left( \mathcal{H}_{ij}^{\kappa \alpha} R_{ij}^{\beta} + \mathcal{H}_{ij}^{\kappa \beta} R_{ij}^{\alpha} \right).
 \end{equation}
 In Eq.~\ref{eq:naforce}, $R_{ij}^{\alpha}$ is the $\alpha$ direction of the vector $\mathbf{R}_i - \mathbf{R}_j$.  Note that the force field $\mathbf{\Xi}_a$ is zero in a symmetric 
 environment, and thus is zero for most crystals. Therefore, we obtain the expect result that there is zero sound attenuation for harmonic crystals. 
 
 For a pair potential $V_{ij}\equiv V(R_{ij})$ Eq.~\ref{eq:naforce} can be written as
 \begin{equation}\label{eq:napair}
 \Xi_{\alpha \beta,i}^\kappa = - \sum_j \left(R_{ij} \frac{\partial^2 V_{ij}}{\partial R_{ij}^2} - \frac{\partial V_{ij}}{\partial R_{ij}} \right)
 \frac{R_{ij}^\kappa}{R_{ij}} \frac{R_{ij}^\alpha}{R_{ij}}\frac{R_{ij}^\beta}{R_{ij}}.
 \end{equation}
 We will refer to this form later to make connections with other work. 
 
\begin{figure}
 \includegraphics[width=0.9\columnwidth,bb=0in 0in 8.5in 8in]{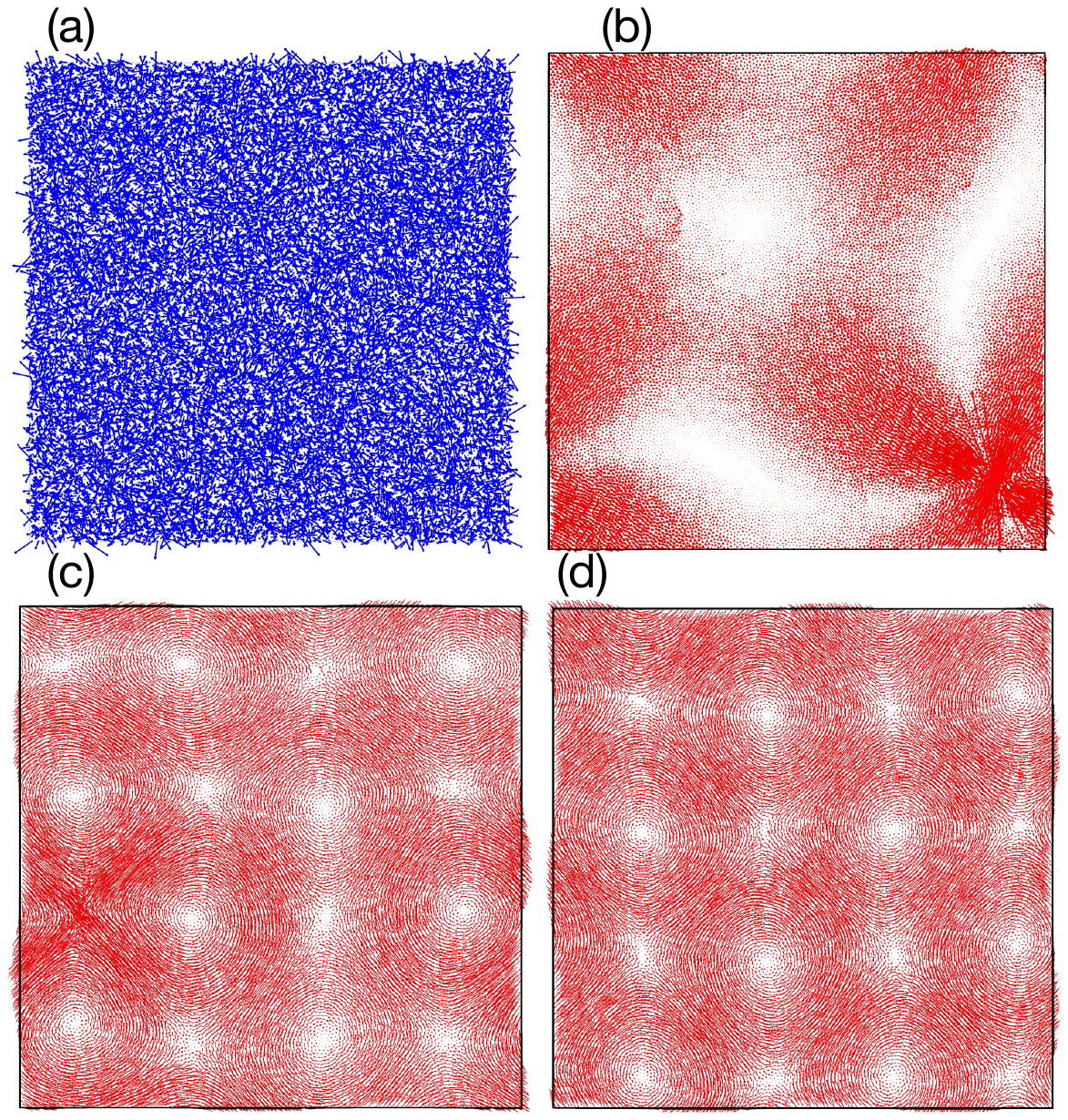}
 \caption{\label{fig:naf} Examples of two vector fields required for low-frequency sound attenuation calculations. (a) The non-affine force field, where the force field is shown
 for a simple shear deformation. (b) An eigenvector of the Hessian where a quasi-localized excitation is clearly seen in the 
 lower right corner. (c) An eigenvector of the Hessian corresponding to a linear combination of plane waves and a hybridized 
 quasi-localized excitation. (d) An eigenvector of the Hessian where no quasi-localized excitations are present, and the 
 linear combination of plane waves is similar to what is seen in (c). Sound attenuation is related to the projection of these two vector fields.}
 \end{figure}

 Shown in Fig.~\ref{fig:naf}(a) is a non-affine force field for a simple shear deformation for a glass. At first glance the forces look uncorrelated, but by Newton's third law
 the forces have to be correlated to at least the range of the interaction potentials \cite{Keta2023}.
 
 The second component of calculating sound attenuation are the eigenvectors $\boldsymbol{\mathcal{E}}(\omega_p)$ where the frequency of the sound wave $\omega \approx \omega_p$. 
 For a large enough finite sized simulations, the eigenvalues are found in bands nearly centered around $\omega = c_a q$ where $c_a$ is the speed of sound. 
 
 One can use the participation ratio $p(\omega_p)$ to get an idea of the vibrational mode is localized at just a few atoms or if the vibrational mode is extended. Plane waves 
  have participation ratios on the order of unity, and a mode localized at one atom has a participation ratio of $1/N$. If the frequency of the vibrational mode is not
  close to $c_\alpha |\mathbf{q}|$, where $|\mathbf{q}|$ is an allowed wavevector due to periodic boundary conditions, and the participation ratio is small, then the vibrational 
  mode often resembles the mode shown in Fig.~\ref{fig:naf}(b). These modes are typically associated with quasi-localized excitations 
  that are not strongly hybridized with plane waves. 
  
 Shown in Fig.~\ref{fig:naf}(c) is an eigenvector for a stable two-dimensional glass. For eigenvectors in this frequency range it is common to find 16 areas where the components 
 are small, which would correspond to 16 white spots in the figure, see Fig.~\ref{fig:naf}(d). 
 This is due to the eigenvector being a linear combination of plane waves. Here we only find 15 areas where the components
 of the eigenvector are small in Fig.~\ref{fig:naf}(c). Where we would expect another area the components of the eigenvector corresponding to the individual  particles are rather large and, 
 visually, appear to be similar to what has been identified as a 
 quasi-localized excitation as shown in Fig.~\ref{fig:naf}(b) \cite{Gartner2016,Wijtmans2017,Kapteijns2020,Lerner2018}. 
 
  We note that a participation ratio of exactly $1/N$ often indicates a particle that is not fully constrained, i.e.~a rattler. In practice, it has always been found that these eigenvectors
  are at a low frequency far away from any sound wave frequency and have a small enough contribution to sound attenuation that they can be ignored. 
  
  In previous work we examined $\epsilon(\omega_p)$ and found that that the eigenvectors with the largest contribution to sound attenuation also had non-uniform magnitudes of the
  components corresponding to different particles \cite{Flenner2024}. 
  By examining the size of the components of the eigenvectors $|\boldsymbol{\mathcal{E}}_i|$ corresponding to particle $i$ over a range of frequencies, 
  we found that there are regions in space that influence low-frequency sound 
  attenuation more than other regions in space. 
  
\begin{figure}
    \includegraphics[width=0.9\columnwidth,bb=0in 0in 8.5in 8in]{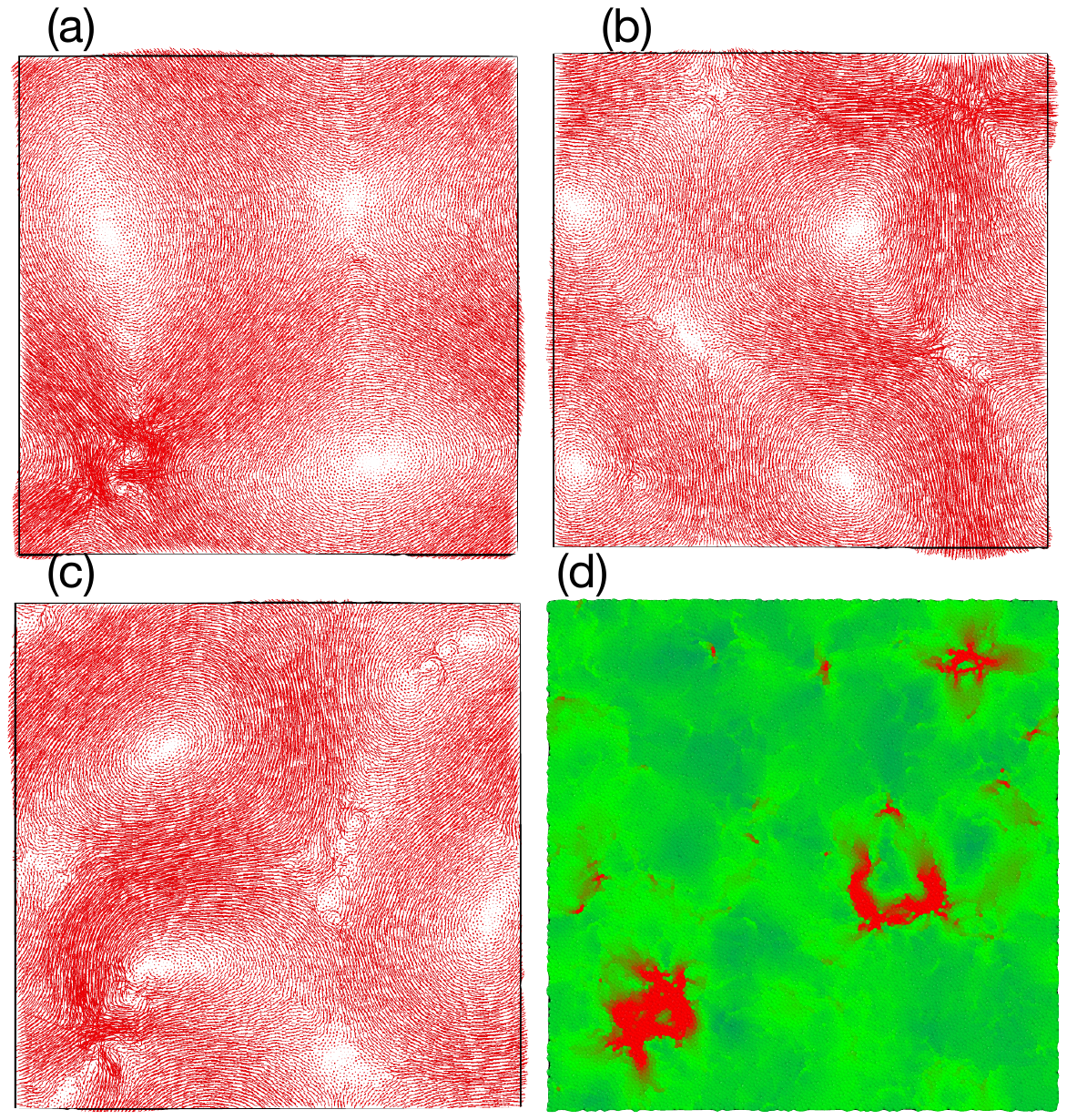}
  \caption{\label{evdefect}Three eigenvectors of one glass configuration for a poorly annealed two-dimensional glass (a)-(c). There is a region in the lower left hand corner in 
  (a) and (c) where the magnitude of the components of the eigenvectors are noticeably larger than in other regions. In (b) we can see regions where the components are also larger.
  In (d) the red patches are the attenuation defects determined from the 24 lowest frequency eigenvectors for the configuration shown in (a)-(c). }
  \end{figure}

  As an illustration, we show three eigenvectors for the same configuration of the two-dimensional glass previously studied, Fig.~\ref{evdefect}(a)-(c).
  We can see that the components of the eigenvectors corresponding to individual particles are large in one specific region of space, which leads us to believe that there is some structural 
  origin. Using this observation, in Ref. \citenum{Flenner2024} we defined regions of space whose eigenvector components are, on average, larger than what is expected for a plane wave. 
  These regions, which we referred to as defects, are  shown in red in Fig.~\ref{evdefect}(d) for the configuration shown in Fig.~\ref{evdefect}. We note that these
  defects are not associated with a single vibrational mode, but their structure is often similar to that of quasi-localized modes that are defined using single vibrational modes.
  This observation suggests that quasi-localized modes produce strong sound attenuation. 
  
  However, since the non-affine force field is non-zero for any non centro-symmetric structure, our theory predicts sound attenuation even if the eigenvector is a plane wave. Additionally,
  by approximating the eigenvectors as plane waves, for very stable two-dimensional glasses we predict sound attenuation to within the error of the calculation\cite{Flenner2024}. 
  We could not
  find any defect regions in these glasses, and thus we conclude that defects and, by implication, quasi-localized excitations may not be needed for sound attenuation in amorphous solids. 
  
  If a defect and thus a quasi-localized excitation does exist, it is a strong source of sound attenuation. 
  This leads to the unanswered question, what is the relative role of quasi-localized excitations 
  in experimental glasses? Do they dominate the Rayleigh regime of sound attenuation as well as the tunneling regime? 
  
  The above examination of defects focuses on the structure of the eigenvectors, but the non-affine forces are also important in the calculation of sound attenuation. We note that 
  these forces also enter into the calculation of the correction to the Born term for the elastic constants \cite{Lemaitre2006}. 
  A connection between the elasticity, non-affine forces, and sound attenuation is implied by this observation but it has not yet been fully explored yet. It is possible that a description 
  on the influence of the non-affine forces on amorphous solids' elasticity can also address sound attenuation. We believe that this would be in the spirit of fluctuating elasticity theory. 
  
  In this section we briefly reviewed previous results of our successful theory of sound attenuation. Previous approaches have also 
  shed light onto the sound attenuation, and a physical picture is greatly enhanced by critically examining these approaches. This allows us to put our theory into broader context. 
  
  \section{Defects and Quasi-Localized Excitations} 
 
 The soft potential model addresses the experimental observations of the low temperature behavior of heat capacity, thermal conductivity, and sound absorption
 from below 1K to around 10K - 20K \cite{Ramos2023,Zaitlin1975,Buchenau1972,Schober2011}. The model reproduces the results predicted by the two-level
 tunneling states model \cite{Ramos2023}, and thus accounts for the behavior below 1K.

 The soft potential model assumes that in amorphous solids there exist soft modes that couple
 bi-linearly with the strain field of longitudinal and transverse sound waves, with coupling constant $\Lambda_a$, where $a=L, T$.  The coupling constants are often shown to 
 be related and $\Lambda_L^2 \approx 3 \Lambda_T^2$ \cite{Ramos2023}. The coupling between soft modes and sound waves leads to additional low frequency vibrational modes
 that account for the behavior above 1K.  
 
 As far as we know, there has only been one attempt to determine the bi-linear coupling between sound waves and soft excitations \cite{Buchenau2021}, but the results have not 
 been directly compared to sound attenuation in simulations. 
 The soft-potential model has been used to fit the thermal conductivity from below 1K to above 10K 
 for several glasses \cite{Ramos2023}.
 Future work should directly compare the predictions of the soft potential model, without the inclusion of quantum effects, to simulated glasses. 
 We note, however, that the soft potential model predicts that the frequency scaling of sound attenuation matches the frequency scaling of quasi-localized excitations, and
 that this scaling is $\omega^4$ independent of dimension. 
 
 In contrast, it has been found that low frequency sound attenuation scales as $\omega^3$ in two dimensions \cite{Flenner2024}. In addition,
 the frequency scaling of quasi-localized excitations in two dimensions has been debated \cite{Wang2021,Lerner2022}. 
 These two observations, and the additional ease of the calculations of the bi-linear coupling in 
 two dimensions, makes two-dimensional systems attractive candidates for a detailed comparison of the soft-potential model with computer simulations.
 
 A strong influence of quasi-localized excitations on sound attenuation is likely even if the current formulation of the soft potential model is shown to need modification. 
 Previous works provide indirect evidence that quasi-localized excitations strongly influence sound attenuation. Lerner and Bouchbinder \cite{Lerner2018} showed that
 internal stresses, \textit{i.e.} non-balanced pair forces at the inherent structure positions, are responsible for quasi-localized excitations.
 In turn, Kapteijns \textit{et al.}~showed that by removing these internal stresses sound attenuation was greatly reduced \cite{Kapteijns2021}.  
 To investigate the role of the internal stresses the authors of Ref. \citenum{Kapteijns2021} re-wrote the definition of the Hessian, Eq.~\ref{eq:H}, separating terms
 proportional to the 2nd and 1st derivative of the pair potential with respect to the interparticle distance, and then they decreased the size of the first derivative term.
 Upon examining Eq.~\ref{eq:napair}, one sees that this 
 also changes the non-affine forces.  Therefore, in their study both components that we use to calculate sound attenuation were being changed. We do expect, however, that
 the dramatic decrease in sound attenuation that Kapteijns \textit{et al.} observed was mainly due to the reduction of quasi-localized excitations. 
  
 While previous work did 
 provide evidence that removing quasi-localized excitations play a large role in sound attenuation, it remained unclear whether removing the internal stresses in the 
 Hessian effectively changed the interactions and if this was also a source of reduced sound attenuation. Our work verified the hypothesis that the quasi-localized excitations
 where a large contribution to sound attenuation. Removing these quasi-localized excitations, which we achieved by studying sound attenuation in computer-generated very stable glasses,
 greatly reduces sound attenuation.
 
 There are also two other interesting results that can be inferred from the above mentioned work\cite{Kapteijns2021}. First, as we have already stated above, we suspect that due to the large decrease
 of internal stresses the glasses with the smallest sound attenuation probably had no quasi-localized excitations. However, these glasses had finite sound attenuation. While not directly stated in 
 Ref. \citenum{Kapteijns2021}, this is evidence that defects, at least in the form of quasi-localized excitations, are not a necessary condition for sound attenuation.
 In our work \cite{Flenner2024} we extrapolated the dependence of the density of the sound attenuation defects, that we defined, on the glass stability and we concluded that  
 sound attenuation can occur without defects. The overall conclusion here is that defects, and quasi-localized excitations,
 can strongly influence sound attenuation but they are not necessary for sound attenuation. 
  
 The second additional interesting result of Ref. \citenum{Kapteijns2021}
 is that the sample-to-sample fluctuations of the elastic constants combined with the fluctuating elasticity theory predict the relative variation of sound attenuation.
 We discuss the fluctuating elasticity theory and this prediction in the next section. 
  
 Finally, we wanted to mention very interesting contribution of Mahajan and Pica Ciamarra, who examined the dynamics of an excited sound wave in time and space \cite{Mahajan2024}.
 They found that sound attenuation starts in localized regions
 that corresponded to the centers of the quasi-localized vibrational modes. At later times sound attenuation apparently evades the rest of the glass diffusively. 
 An interesting follow up study would be to examine sound attenuation in the very stable glasses where we cannot identify any defects or quasi-localized excitations. 
  
  \section{Fluctuating Elasticity Theory}
  
  There is a large body of work devoted to connection between fluctuating elasticity and the boson peak
  \cite{Ramos2023,Kapteijns2021,Schirmacher2010,Schirmacher2011,Marruzzo2013,Ganter2010}. 
  In this perspective we focus on a different aspect of the fluctuating elasticity theory, \textit{i.e.} its description of sound attenuation.
  An early study \cite{Marruzzo2013} claimed very good agreement between the theory and simulations. More recently, predictions of a microscopic version
  of the fluctuating elasticity theory were compared with direct simulations of sound attenuation \cite{Lemaitre2019}. 
  The conclusion was that fluctuating elasticity under predicts sound attenuation by about two orders of magnitude.  However, as briefly mentioned in the penultimate paragraph
  of the last section, there is evidence that the change of the elastic constants fluctuations computed in one specific way correlates with the relative change of sound attenuation. 
  
  Motivated by the observation of Kapteijns \textit{et al.}\cite{Kapteijns2021},
  we examined the stability dependence of the sample-to-sample variation of the shear modulus for a model two-dimensional glass
  former and a model three-dimensional glass former. As we mentioned earlier, we label stabilities of different zero-temperature glasses by the temperatures which the liquid
  was equilibrated before it was quenched to the nearest inherent structure. This temperature is called the parent temperature $T_p$. 
  Our goal was to determine whether the change of  the sample-to-sample fluctuations correlates with the change of sound attenuation, in other words whether the sound attenuation
  coefficient is proportional to mean-squared sample-to-sample shear modulus fluctuations. To check this, we fixed the coefficient of proportionality between
  the sound attenuation and the mean-squared sample-to-sample shear modulus fluctuations to reproduce quantitatively the latter quantity for $T_p=0.2$, for both
  two-dimensional and three-dimensional systems. 
  
\begin{figure}
  \includegraphics[width=0.95\columnwidth]{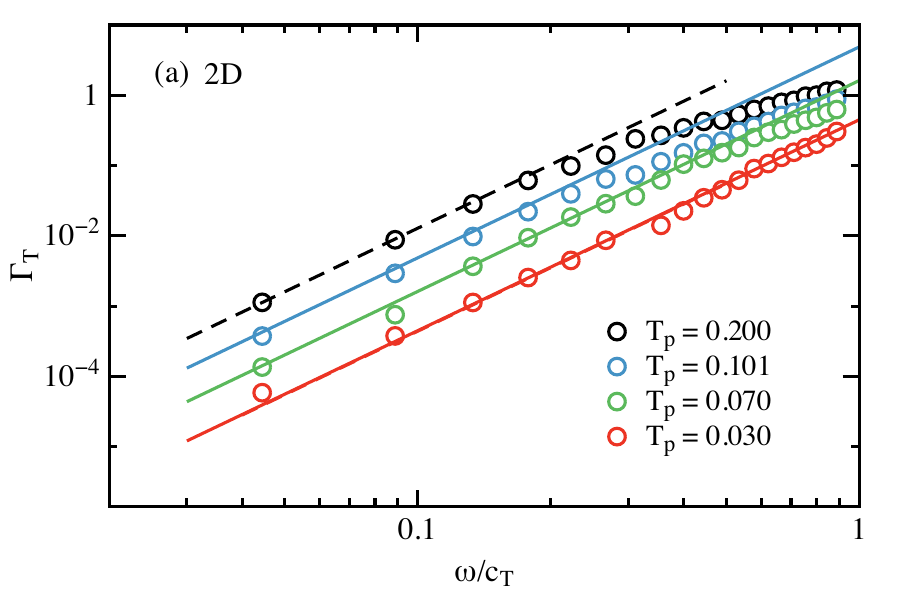}
  \includegraphics[width=0.95\columnwidth]{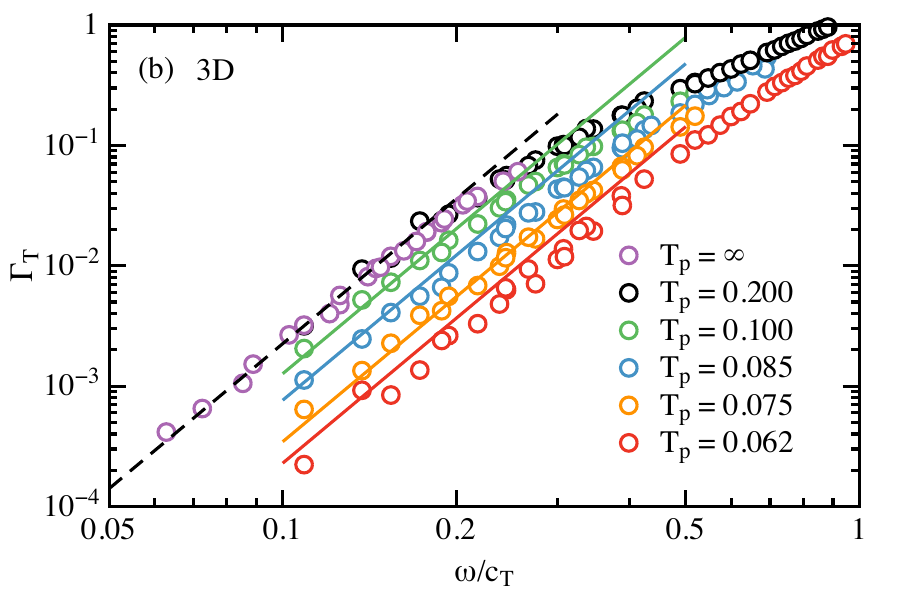}
  \caption{\label{fig:atten} Comparison of sound attenuation coefficients calculated from
    direct simulations with predictions of the fluctuating elasticity theory. Open circles: results of direct sound attenuation simulations \cite{Wang2019a,Flenner2024}.
    Lines: predictions of the fluctuating elasticity theory. The overall magnitude of the theoretical prediction is fixed by fitting theory against the simulation at $T_p = 0.2$.
    $T_\text{p}$ is the parent temperature of the glass \cite{Wang2019a,Wang2019};
    glass stability increases with decreasing parent temperature.}
  \end{figure}

  Shown in Fig.~\ref{fig:atten} is the sound attenuation coefficient calculated from simulations (symbols) and the sound attenuation inferred from fluctuating elasticity (lines) for 
  (a) the two-dimensional system and (b) the three dimensional system. We thus confirm the observation first made by Kapteijns \textit{et al.} that the relative change of sound attenuation
  is accurately predicted by the shear modulus mean-squared sample-to-sample fluctuations in two dimensions
    and extend it to three dimensions. We note that the correlation works slightly better in two dimensions than in three dimensions. 

  We emphasize that for the above described correlation one needs to use sample-to-sample fluctuations of the shear modulus. Our previous work \cite{Wang2019a} suggested that
  the fluctuating elasticity theory did not correctly predict the relative change of the sound attenuation. This conclusion was based on the calculation that used fluctuations of locally
  defined shear modulus, \textit{i.e.} of shear modulus associated with small sub-volumes of the simulation box  \cite{Shakerpoor2020}. We note that there is some arbitrariness in
  the definition of the shear modulus associated with a sub-volume of the whole sample. Mizuno \textit{et al.} \cite{Mizuno2013} showed that different definitions of the local
  shear modulus lead to somewhat different results for its fluctuations. We feel that it would be worthwhile to investigate in future work whether sample-to-sample fluctuations
  are the same as fluctuations of the local modulus defined in a specific way. 
  
  We note that we did not assume any spatial correlations between the fluctuations of the elastic constants. 
  However, a study by Mahajan and Pica Ciamarra suggests that correlated fluctuating elasticity theory is needed to describe
  the relative variation of sound attenuation \cite{Mahajan2021}.
  These authors examined the correlation between sound attenuation evaluated in direct simulations and elastic constants fluctuations for two systems. One of these systems was 
  the system we studied, see results shown in Fig.~\ref{fig:atten}(b). Correlated fluctuating elasticity theory relates the boson peak position and sound attenuation
  to the mean-squared fluctuations of the elastic constants. By relating the change of the position of the boson peak with the change of the mean-squared fluctuations,
  Mahajan and Pica Ciamarra were able to predict 
  the change of sound attenuation for the more stable glasses. However, their study was more restricted than what we showed in Fig.~\ref{fig:atten}
  in that they did not include the least stable glass in the analysis.

  We shall add that Mahajan and Pica Ciammarra attempted to bridge the gap between defect-based approach and fluctuating elasticity theory. To this end they envisioned
  the glass as an elastic medium with defects being spatial regions with elastic constants different from that of the surrounding medium. They estimated of the defect number
  density and suggested that it might be a universal constant. This suggestion contrasts with our finding that the density of particles that are parts of defects defined in Refs.
  \citenum{Flenner2024,Flenner2024b} depends on the parent temperature of the glass, and thus on the glass stability.

  We recall that the existence and nature of spatial
  correlations of local elastic constants is somewhat controversial with different earlier works leading to
  different conclusions \cite{GelinNatMat2016,Mizuno2018,Shakerpoor2020}. This issue should be investigated in future work.

  It would be worthwhile to develop a first principles fluctuating elasticity theory of sound attenuation and to test it against computer simulations on glasses of widely different
  stability.  We hope that our description of sound attenuation in terms of non-affine forces and properties of eigenvectors can aid in the formulation of this theory.
  Such a theory should be able to account for quasi-localized excitations and defects. 
  
  \section{Euclidean Random Matrix model}
  
  As we discussed in Sec.~\ref{nonaff}, sound attenuation is both detected in direct computer simulations and predicted by the theory in glassy samples
  without any regions of space with unusually large components of the low-frequency eigenvectors, \textit{i.e.} without defects. 
  This finding agrees with results presented in Bouchbinder and Lerner's works that suggested that sound attenuation can occur in samples without quasi-localized excitations. 
  In this section we review a simple alternative model that exhibits the same behavior.

  Euclidean Random Matrix (ERM) model was introduced by M\'{e}zard, Parisi and Zee \cite{Mezard1999} as a toy model for vibrational spectra of amorphous systems.
  The original publication focused on a bit artificial version of this model for which the uniform displacement is not a zero eigenvalue eigenvector. Subsequent works
  \cite{Martin2001,Grigera2001,Ciliberti2003,Grigera2011,Ganter2011,Schirmacher2019,Conyuh2021,Vogel2023,Baumgartel2024,Szamel2025} analyzed a more realistic version
  of the model that postulated a euclidean random matrix that was positively definite and for which the  uniform displacement is an eigenvector corresponding to a
  zero eigenvalue. Most of these studies used the so-called scalar version of the ERM model, in which the vector character of the interparticle forces and displacements is ignored.
  The most realistic version, in which the displacement field is a vector field, was studied by Ciliberti \textit{et al.} \cite{Ciliberti2003}.

  The scalar version of the ERM model starts with a matrix $\mathbf{M}_{ij} = f(\mathbf{R}_{ij})-\delta_{ij}\sum_k f(\mathbf{R}_{ik})$, with $\mathbf{R}_{ij}=\mathbf{R}_i-\mathbf{R}_j$,
  where quenched positions $\mathbf{R}_i$ are distributed independently and randomly, and $f(\mathbf{R})$ is referred to as spring function.
  Matrix $\mathbf{M}$ plays the role of the Hessian matrix of the glass.
  The advantage of this model is that it allows one to derive a perturbative expansion in inverse number density of the quenched positions.
  After some initial confusion it is now well established that the ERM model predicts Rayleigh scattering scaling of sound attenuation \cite{Grigera2011,Ganter2011}. 
    
  Recently Baumg\"artel, Vogel, and Fuchs\cite{Baumgartel2024} studied an ERM model where the spring function is isotropic and given by a Gaussian function.
  The results of this study qualitatively agree with the results of our analysis of vibrational properties and sound attenuation of very stable glassy samples without defects.
  
  First, Baumg\"artel \textit{et al.} did not find any low frequency localized modes: it is evident from Fig.~11 of Ref. \citenum{Baumgartel2024} that the
  participation ratios of the low frequency modes are as expected for plane waves. This is perhaps not surprising due to the completely random (Poisson) distribution of
  the quenched positions. 

  Additionally,  Baumg\"artel \textit{et al.} found that the number of modes in each frequency band is given by what is  expected from the 
  Debye theory\cite{BVFprivate}. However, they also found that in ERM there is still a boson peak and there are low frequency modes in excess of the Debye theory.
  The excess low frequency modes are due to the wavevector, i.e. frequency, dependence of the speed of sound increasing the number of modes at a fixed frequency to be above
  that of the frequency independent speed of sound assumed in Debye theory.  There are still excess modes over Debye theory without quasi-localized modes. 
  
  Second, there is Rayleigh scattering scaling of sound attenuation at low frequencies. However, the magnitude of the sound attenuation is very small.
  This observation provides additional evidence that defects and associated quasi-localized excitations are not necessary for Rayleigh scattering scaling.

  Future work should extend recent self-consistent theories for sound propagation and attenuation in ERM models \cite{Vogel2023,Szamel2025} to the
  vector version of the ERM. In particular, such an extension will allow to investigate the influence of internal stresses, \textit{i.e.} un-balanced pair forces,
  on sound attenuation. 

  In addition, within the ERM model one could examine fluctuations of local elastic constants, \textit{e.g.} defined as in the supplementary information of Ref. \citenum{Mahajan2021}, 
  and test if these fluctuations predict the density dependence of sound attenuation. 
  If so, there would be compelling evidence that there must be some way to formulate fluctuating elasticity theory to predict sound attenuation with no adjustable parameters. 
  
  \section{Conclusions}

  In the past decade sound attenuation has been recognized as a useful and informative window into low temperature properties of glasses. The research in this area was
  facilitated by new computational methods that allowed researchers to create computer-generated glasses of widely varying stability. This facilitated investigations
  of glass samples with the same interactions but very different properties. Thus, existing sound propagation theories could be used to predict stability dependence of
  sound attenuation and these predictions could then be tested against results of direct simulations. We hope that the resulting progress in our understanding of sound attenuation will
  significantly enhance our knowledge of the low temperature properties of glasses.
  
  Recent work on the importance of non-affine forces represents a new direction in our understanding of sound attenuation.
  It is clear, however, that previous approaches at least provide a piece of the physical interpretation. There is ample evidence 
  that some quasi-localized excitations play a significant role in sound attenuation when they are present. However, at the same time there is mounting evidence that quasi-localized excitations are 
  not necessary for sound attenuation. An outstanding question is whether these two contributions can be clearly separated and understood individually. 
  
  Additionally, computer simulations show that fluctuating elasticity theory can, at least, predict the relative change of a glasses stability with stability or other control parameters. 
  This observation suggests that we should look at the role of quasi-localized excitations in fluctuating elasticity. Additionally, more theoretical effort should be extended to
  a first principles derivation of fluctuating elasticity that may allow for a first principles prediction without any adjustable parameters, as is done with current comparisons with
  simulations. 

  An interesting feature that has emerged in recent years is Rayleigh scaling of sound attenuation in systems without low-frequency quasi-localized excitations, \textit{i.e.} very stable 
  two-dimensional glasses and Euclidean Random Matrix-based glass models. It is now clear that the presence of well defined isolated defects is not necessary for Rayleigh scattering scaling
  to occur. We argued in Ref. \citenum{Flenner2024} that at low frequencies sound attenuation is proportional to $\omega^2 g(\omega)$ where $g(\omega)$ is the density of states.
  This relationship is referred to as the ``Rayleigh-Klemens'' law and was derived earlier using a variety of methods \cite{Moriel2019,Maurer2004,Wyart2010,DeGiuli2014}.
  Since the dominant contribution to the density of states is given by the Debye theory\cite{Mizuno2017},  this relation predicts that the dominant contribution to sound attenuation
  exhibit Rayleigh scattering scaling. There should also be a correction that is proportional to the density of excess modes. Current simulations are not sensitive enough to 
  detect this small correction.

\section*{Author declarations}

\subsection*{Conflict of interest}

The authors have no conflicts to disclose.

\section*{Data availability}

The data that support the findings of this study are available from
the corresponding author upon reasonable request.
 
  \section*{Acknowledgments}

  We thank E. Bouchbinder for a useful comment on the manuscript. We gratefully acknowledge the support of NSF Grant No. CHE 2154241.

  \end{document}